\documentclass[english]{article}
\usepackage[T1]{fontenc}
\usepackage[latin9]{inputenc}
\usepackage{color}
\usepackage{textcomp}
\usepackage{mathrsfs}
\usepackage{amsmath}
\usepackage{amssymb}
\usepackage{esint}

\makeatletter

\newcommand{\noun}[1]{\textsc{#1}}

\newcommand{\lyxaddress}[1]{
	\par {\raggedright #1
	\vspace{1.4em}
	\noindent\par}
}

\makeatother

\usepackage{babel}
\begin{document}
\title{Role of the electromagnetic vacuum in the transition from classical
to quantum mechanics}
\author{Ana María Cetto and Luis de la Peña}
\maketitle

\lyxaddress{Instituto de Física, Universidad Nacional Autónoma de México, 04360
Mexico City, Mexico}
\begin{abstract}
We revisit the nonrelativistic problem of a bound, charged particle
subject to the random zero-point radiation field (\noun{zpf}), with
the purpose of revealing the mechanism that takes it from the initially
classical description to the final quantum-mechanical one. The combined
effect of the \noun{zpf} and the radiation reaction force results,
after a characteristic time lapse, in the loss of the initial conditions
and the concomitant irreversible transition of the dynamics to a stationary
regime controlled by the field. In this regime, the canonical variables
$x,p$ become expressed in terms of the dipolar response functions
to a set of field\noun{ }modes. A proper ordering of the response
coefficients leads to the matrix representation of quantum mechanics,
as was proposed in the early days of the theory, and to the basic
commutator $\left[\hat{x},\hat{p}\right]=i\hbar$. Further, the connection
with the corresponding Fokker-Planck equation valid in the Markov
approximation, allows to obtain the (nonrelativistic) radiative corrections
of \noun{qed}. These results reaffirm the essentially electrodynamic
and stochastic nature of the quantum phenomenon, as proposed by stochastic
electrodynamics.
\end{abstract}

\section{Introduction}

This paper is devoted to an analysis of the process that takes the
system described by stochastic electrodynamics \noun{(sed)}---the
combined system composed typically of a charged subatomic particle
subject to an external potential plus the zero-point radiation field
(\noun{zpf})---from an initially classical deterministic behavior
to its final behavior, described by quantum mechanics (\noun{qm}). 

The analysis presented is in line with the notion that has guided
the efforts of a number of authors working on \noun{sed}, namely that
this theory can serve as a legitimate underpinning for \noun{qm}.
This implies that \noun{sed} and \noun{qm} can be simultaneously valid
only by accepting the transition from classical to quantum as a real
phenomenon. Up to now the existence of such process has been intuitively
anticipated (although not always recognized), but no detailed satisfactory
description of the transition exists so far.

One widespread line of thought within the \noun{sed} community has
been to consider the action of the \noun{zpf} as a perturbation to
an otherwise classical motion, with the expectation that the quantum
properties emerge as a consequence. Although this perturbative approach
has led to a series of positive results (\cite{Marshall result}-\cite{TQD}
and further references therein), it has also met with apparently unsurmountable
difficulties, such as self-ionization of the H atom \cite{Cole result,Theo H},
and more generally it has not been able to reproduce the quantum formalism,
even as an approximation \cite{PertSED,HuangBat}. 

In this connection it is worth recalling Nelson's stochastic mechanics
\cite{Nelson66} (see also \cite{LP69}, \cite{TQD} Ch. 2 for an
alternative version; we refer to either of these theories as stochastic
quantum mechanic\noun{s, sqm}), which is a phenomenological stochastic
formulation leading successfully to \noun{qm}. It may be easily realized
that \noun{sqm} is \textit{\emph{not}}\emph{ }a classical theory,
since its Newton's law of motion violates the classical deterministic
canons by the addition of a stochastic acceleration as part of its
mechanical description, and by the appearance of a diffusive velocity
$\boldsymbol{u}(\boldsymbol{x},t)$, from which derives the quantum
potential $V_{Q}=\frac{1}{2}(m\boldsymbol{u}^{2}+\hbar\nabla\cdot\boldsymbol{u})$.
Thus, \noun{sqm} does not imply a \emph{transition} from classical
to quantum: it is already \emph{by construction} a non-classical theory.

As extensively discussed in \cite{TEQ} and references therein, a
non-perturbative statistical treatment of the \noun{sed} problem reproduces
in the time-asymptotic limit the quantum formalism; we therefore refer
to the regime attained in this limit as the \emph{quantum regime}.
Further, it has been demonstrated more recently \cite{Nos 20 o 21}
that in the quantum regime, the \noun{sed} mean equations of motion
coincide in every detail with those of \noun{sqm}, which indicates
that in this regime the two descriptions are equivalent in their dynamical
content. Both include the pair of mean local velocities, the flow
velocity $\boldsymbol{v}$ and the diffusive velocity $\boldsymbol{u}$,
which are also contained---even if not so conspicuously---in \noun{qm}.
Although one might therefore consider that the three are basically
the same non-classical theory, their principles differ substantially.
To arrive at \noun{qm} \textit{from a classical starting point}---i.
e., without introducing quantum postulates---an ontological element
must be added to resolve the underdetermination of \noun{qm} \cite{Egg21,Cush94-1},
one that accounts for both the indeterministic and the electromagnetic
nature of the quantum phenomenon, as will become clear below. \noun{sed}
posits the random \noun{zpf} as this ontological element, an ingredient
that is conspicuously absent in classical physics and alas hidden
in \noun{qm}. But as it turns out, the \noun{sed} approach leading
consistently to \noun{qm} is not merely perturbative; it must take
into account the essential role of the \noun{zpf}.

The motivation of the present paper is precisely to discuss the transition
to \noun{qm} based on the observation that the \noun{zpf,} in combination
with the radiation reaction force, plays a crucial role in irreversibly
modifying the dynamical behavior of the particle. As an end effect,
the variables required for the (apparently) mechanical description
in the quantum regime carry with them an indelible imprint of the
background field, which becomes concealed in the quantum formalism.

The approach followed here is not entirely new; a first proposal along
this direction was advanced in \cite{nos1991} and a more elaborate
version was presented in \cite{TQD}, Ch. 10. More recent work that
serves as background to the present paper is contained in Refs. \cite{TEQ,QSMF20}.
In the interest of clarity, some previously published results are
included here; however, we omit in general the detailed derivations
and focus the discussion on the key elements that, taken together,
help shape a coherent picture of the process leading to quantum mechanics.
Herein lies the originality of the present paper: it offers a more
integrated causal explanation of the emergence of quantum mechanics
in its matrix formulation, from the perspective of \noun{sed}.

The structure of the paper is as follows. Section 2 contains a brief
qualitative description of the most relevant features of the evolution
of the \noun{sed} system towards the quantum regime, with the purpose
of providing an introduction to the quantitative discussion of section
3. Subsection 3.1 contains an analysis of the influence of the \noun{zpf}
on the kinematics of the particle, carried out within the framework
of Hamiltonian dynamics. This analysis shows that, as a result of
the decay of the deterministic motion, the \noun{zpf} takes control
over the particle response, so that the description in terms of the
initial canonical variables $x,p$, changes to one that depends on
the canonical variables of the field. In subsection 3.2 we identify
the response coefficients with the matrix elements of the operators
$\hat{x},\hat{p}$ that describe dipolar transitions. In subsection
3.3 we follow a statistical approach and use the results of the previous
section to determine the diffusion coefficients appearing in the ensuing
Fokker-Planck equation, valid in the quantum regime. In Subsection
3.4 we describe the time evolution of the average energy in this regime
and derive the \noun{qed} formula for the spontaneous emission coefficient.
The paper ends with a brief discussion of the most important implications
of the results obtained, for a better understanding of the physical
process leading to the quantum formalism.

\section{The \noun{sed} process towards equilibrium\label{TE}}

Let us consider an \noun{sed} system consisting of a charged particle
(typically an electron), which is subject to the action of a conservative
binding force and at some instant (say, at time $t_{o}$) gets connected
to the random \noun{zpf}, which has an energy $\hbar\omega/2$ per
mode. In practice because of the stochastic nature of the problem,
when speaking of one particle one has in mind an ensemble of similar
systems. At the outset the dynamics is essentially that of a classical
electrodynamic system; in particular, the equations of motion are
satisfied with the initial values of the mechanical variables, which
are freely specified. On the other hand, the variables describing
the instantaneous state of the \noun{zpf} are determined by purely
hazardous circumstances, beyond our control and even our knowledge.
Therefore, the initial dynamics follows the classical deterministic
rules of electrodynamics; however, this picture holds for a short
time, as the random \noun{zpf} compels the particle to start moving
unpredictably. The system then begins to accumulate information about
the random motions impressed upon it, by constructing the diffusion
tensors step by step (see Eqs. (\ref{dif tens}) below). At the same
time, the radiation reaction force due to the accelerations forced
upon the particle by the external force and the \noun{zpf,} has a
dissipative effect on the particle dynamics. 

In the end, a remarkable process has taken place: the accumulated
effects of the ran\textcolor{black}{dom wave fi}eld---which conveys
memory through its correlations---have \textit{\emph{generated}}\emph{
}an important diffusive component of the velocity, associated with
the inhomogeneities in the distribution density of particles and giving
rise in its turn to the quantum potential (\cite{TEQ} and references
therein). (Analogous phenomena of density accumulation due to a random
force are discussed in detail in Ref. \cite{Klyatskin 20xx}.) Eventually
the diffusion tensors reach their final form, to serve as a statistical
inventory of the fluctuations taking place in the system. Simultaneously---and
most importantly---the radiation reaction force leads to the loss
of memory of the initial conditions. In the absence of external excitations,
the originally deterministic motion governed by the external force
decays and only the (stable) fluctuating motions maintained by the
\noun{zpf} survive; this is characteristic of the ground state. More
generally, the concurrent evolution of\textcolor{black}{{} radiation
reaction and diffusion tensors leads to a convergence of the mean
values of the radiated and absorbed powers, taking the system to a
stationary state \cite{TEQ}. The classical constants of motion, in
co}rrespondence with the new situation, acquire their---stationary
\textit{and} quantized---final value. Indeed, the system has entered
the quantum regime, and henceforth follows the quantum rules. 

This, in a nutshell, we have learned along the years from the development
of \noun{sed}. All in all it implies an extraordinary transition from
the initially deterministic dynamics to a non-deterministic one controlled
by the \noun{zpf}, which has however not been analyzed in detail so
far. In particular, as anticipated in Ref. \cite{QSMF20}, and further
elaborated in what follows, the permanent action of the \noun{zpf}
turns out to induce a gradual and irreversible change in the \textit{nature}
of the variables that describe the particle dynamics: they eventually
become the particle variables \emph{controlled by the field}. This
feature is seen to be at the core of the operator representation of
quantum mechanics.

\section{Quantitative approach}

\subsection{Kinematics of the SED system; algebraic structure\label{KIN}}

In a classical Hamiltonian system, consisting of a particle subject
to a conservative force $\boldsymbol{f}(\boldsymbol{x})$, the (phase-space)
algebraic structure is determined by the Poisson brackets in terms
of the full set of the particle's canonical variables $\left\{ x_{i};p_{i}\right\} $
(we use a semicolon for the set of canonical variables, to distinguish
it from the Poisson bracket), 
\begin{equation}
\left\{ x_{j}(t),p_{i}(t)\right\} _{xp}=\delta_{ij},\label{K4}
\end{equation}
with $i,j=1,2,3$. However, the complete \noun{sed} system, which
is also a Hamiltonian system, consists of the particle subject to
the conservative force plus the radiation field; therefore the full
set of canonical variables must include those of the (infinite number
of) field modes, 
\begin{equation}
\left\{ q;p\right\} =\left\{ x_{i};p_{i}\right\} +\left\{ q_{\alpha};p_{\alpha}\right\} ,\label{K6}
\end{equation}
with $q_{\alpha},p_{\alpha}$ the canonical variables corresponding
to the mode of the radiation field of (circular) frequency $\omega_{\alpha}$.
A discrete set of frequencies is considered here, for reasons that
will become clear later.

At the initial time $t_{o}$, when particle and field start to interact,
the full set of canonical variables is given by

\begin{equation}
\left\{ q_{o};p_{o}\right\} =\left\{ x_{io};p_{io}\right\} +\left\{ q_{\alpha o};p_{\alpha o}\right\} ,\label{K8}
\end{equation}
with $\text{\ensuremath{x_{io}},\ensuremath{p_{io}}}$ the initial
values of the particle's coordinate and momentum components and $q_{\alpha o},p_{\alpha o}$
the canonical variables corresponding to the original field modes,
which are those of the \noun{zpf} alone. Because the system is Hamiltonian,
the variables at time $t,$ $q=q(t),p=p(t)$, are related to $q_{o},p_{o}$
via a canonical transformation, and therefore the Poisson bracket
of any two functions $f,g$ can be taken indistinctly with respect
to the variables at time $t$ or at time $t_{o}$,
\begin{equation}
\left\{ f,g\right\} _{qp}=\left\{ f,g\right\} _{q_{o}p_{o}}.\label{K10}
\end{equation}

Since we are interested in the description of the dynamics of the
particle, our focus will be on the Poisson bracket of particle variables
$x_{i},p_{j}$, which at any time $t$ satisfies of course
\begin{equation}
\left\{ x_{i}(t),p_{j}(t)\right\} _{qp}=\delta_{ij}.\label{K12}
\end{equation}
From Eq. (\ref{K10}) it follows that this is equal to the Poisson
bracket with respect to the full set of initial variables,
\begin{equation}
\left\{ x_{i}(t),p_{j}(t)\right\} _{q_{o}p_{o}}=\left\{ x_{i}(t),p_{j}(t)\right\} _{x_{o}p_{o}}+\left\{ x_{i}(t),p_{j}(t)\right\} _{q_{\alpha o}p_{\alpha o}}.\label{K14}
\end{equation}

As discussed in Ref. \cite{QSMF20} (and considered again in subsection
\ref{DYN}), a result of the dissipative effect upon the systematic
motion due to the radiation reaction force is that the particle eventually
loses memory of its initial conditions $\boldsymbol{x}_{o},\boldsymbol{p}_{o}$.
Therefore, for times larger than a characteristic dissipation time
$\tau_{d}$, Eq. (\ref{K14}) reduces to
\begin{equation}
\left\{ x_{i}(t),p_{j}(t)\right\} _{q_{o}p_{o}\:\overrightarrow{t>\tau_{d}}}\left\{ x_{i}(t),p_{j}(t)\right\} _{q_{\alpha o}p_{\alpha o}},\label{K16}
\end{equation}
and by combining Eqs. (\ref{K12})--(\ref{K16}) we get the condition
\begin{equation}
\left\{ x_{i}(t),p_{j}(t)\right\} _{q_{\alpha o}p_{\alpha o}}=\delta_{ij}\qquad(t>\tau_{d}).\label{K18}
\end{equation}
An order-of-magnitude calculation of a typical dissipation time gives
\cite{RMF76}
\[
\tau_{d}\approx10^{-11}s,
\]
taken as the time needed for a ``classical'' radiating electron
moving initially in an orbit of Bohr radius to collide with the nucleus
\cite{OD17}.

Let us introduce normal field variables $a_{\alpha},a_{\alpha}^{*}$,
such that 
\begin{equation}
a_{\alpha}=e^{i\phi_{\alpha}},a_{\alpha}^{*}=e^{-i\phi_{\alpha}},\label{K20}
\end{equation}
with $\phi_{\alpha}$ statistically independent, random phases in
$\left(-\pi,\pi\right)$, as corresponds to the \noun{zpf}. Since
the total energy per frequency of the mode $\alpha$ of the \noun{zpf},
taking into account the two polarizations, is $\mathscr{E}(\omega_{\alpha})=\hbar\omega_{\alpha}$,
the transformation between normal field variables and field quadratures
is given by
\begin{equation}
\omega_{\alpha}q_{\alpha}^{o}=\sqrt{\hbar\omega_{\alpha}/2}(a_{\alpha}+a_{\alpha}^{*}),\;p_{\alpha}^{o}=-i\sqrt{\hbar\omega_{\alpha}/2}(a_{\alpha}-a_{\alpha}^{*}).\label{K22}
\end{equation}
These transformation rules applied to the Poisson bracket with respect
to the field variables give
\begin{equation}
i\hbar\left\{ x_{i},p_{j}\right\} _{q_{\alpha o}p_{\alpha o}}=\left\{ x_{i},p_{j}\right\} _{a_{\alpha}a_{\alpha}^{*}}=\left[x_{i},p_{j}\right],\label{K24}
\end{equation}
with the bilinear form $\left[f,g\right]$ defined in general as the
transformed Poisson bracket \cite{QSMF20,TQD}, 
\begin{equation}
\left[f,g\right]\equiv\sum_{\alpha}\left(\frac{\partial f}{\partial a}_{\alpha}\frac{\partial g}{\partial a_{\alpha}^{*}}-\frac{\partial g}{\partial a}_{\alpha}\frac{\partial f}{\partial a_{\alpha}^{*}}\right).\label{K26}
\end{equation}
From Eqs. (\ref{K18}) and (\ref{K24}) we conclude that for times
$t>\tau_{d}$ the bilinear form $\left[x_{i},p_{j}\right]$ must satisfy
the condition
\begin{equation}
\left[x_{i},p_{j}\right]=i\hbar\delta_{ij}.\qquad(t>\tau_{d})\label{K28}
\end{equation}

This most important result indicates that the symplectic relation
between the particle variables $x_{i}$ and $p_{j}$ becomes determined
(for $t>\tau_{d}$) by their functional dependence on the normal \noun{zpf}
variables as is expressed in Eq. (\ref{K26}), with the scale given
by Planck's constant. 

\subsection{Stationary states and transitions; operators}

From Eqs. (\ref{K26}) and (\ref{K28}) it is clear that the variables
$x_{i}$, $p_{j}$ are now driven by a set of field modes, denoted
by $\left\{ \alpha\right\} $. In order to further analyze the implications
of this result, let us for the moment consider that sufficient time
has elapsed for the system to have reached the quantum regime, with
stationary states of motion resulting from the combined effect of
the radiation reaction force and the driving force of the (stationary)\noun{
zpf} on the particle \cite{TEQ}. In subsection \ref{DYN} we come
back to the discussion of how this regime is attained.

In the absence of external radiation, the field is in its ground state,
namely the \noun{zpf}, and correspondingly also the particle is in
its ground state, characterized by its energy $\mathcal{E}_{o}$.
In the presence of external excitations the particle may reach an
excited state $n$, with energy $\mathcal{E}_{n}\mathcal{>E}_{o}$.
Of importance in the present context is that the symplectic structure
of the problem is already defined, which means that (\ref{K28}) can
be applied to any stationary state, be it the ground state or an excited
state $n$. Let us tag the corresponding variables $x_{i},p_{j}$
with the subindex $n$, and write, using (\ref{K26}),
\begin{equation}
\left[x_{i},p_{j}\right]_{nn}=\sum_{\alpha}\left(\frac{\partial x_{n}}{\partial a_{\alpha}}\frac{\partial p_{n}}{\partial a_{\alpha}^{*}}-\frac{\partial p_{n}}{\partial a_{\alpha}}\frac{\partial x_{n}}{\partial a_{\alpha}^{*}}\right)=i\hbar\delta_{ij}.\label{T2}
\end{equation}
For ease of notation we refer in what follows to the one-dimensional
case. Equation (\ref{T2}) implies necessarily that the variables
$x_{n}$ and $p_{n}=m\dot{x}_{n}$ are linear functions of $\left\{ a_{\alpha},a_{\alpha}^{*}\right\} $.
The field modes involved are those to which the particle responds,
i. e., those that can take the particle from state $n$ to another
state, say $k.$ We therefore write (\cite{QSMF20}; see also \cite{nos1991,TQD})
\begin{equation}
x_{n}(t)=\sum_{k}x_{nk}a_{nk}e^{-i\omega_{kn}t}\mathrm{+c.c.,\:\;}p_{n}(t)=\sum_{k}p_{nk}a_{nk}e^{-i\omega_{kn}t}\mathrm{+c.c.,}\label{T4}
\end{equation}
where $a_{nk}$ is the normal variable associated with the field mode
that connects state $n$ with state $k$, and $x_{nk}$, $p_{nk}=-im\omega_{kn}x_{nk}$
are the corresponding response coefficients \cite{QSMF20}. Introduction
of these expressions into the bilinear form Eq. (\ref{T2}) gives
\[
\left[x,p\right]_{nn}=2im\sum_{k}\omega_{kn}\left|x_{nk}\right|^{2}=i\hbar,
\]
whence
\begin{equation}
\sum_{k}\omega_{kn}\left|x_{nk}\right|^{2}=\hbar/2m.\label{T6}
\end{equation}
Further, since the field variables $a_{nk},a_{n'k}$ connecting different
states $n,n'$ with state $k$ are independent random variables, by
combining Eqs. (\ref{K26}), (\ref{T4}) and (\ref{T6}) one gets
\begin{equation}
\left[x,p\right]_{nn'}=i\hbar\delta_{nn'}.\label{T8}
\end{equation}
The coefficients $x_{nk}$ and the normal variables $a_{nk}$ refer
to the transition $n\rightarrow k$ involving the frequency $\omega_{kn}$,
whilst $x_{kn}$ and $a_{kn}$ refer to the inverse transition, with
$\omega_{nk}=-\omega_{kn}$; therefore, from (\ref{T4}),
\begin{equation}
x_{nk}^{*}(\omega_{nk})=x_{kn}(\omega_{kn}),\ p_{nk}^{*}(\omega_{nk})=p_{kn}(\omega_{kn}),\ a_{nk}^{*}(\omega_{nk})=a_{kn}(\omega_{kn}),\label{T10}
\end{equation}
whence Eq. (\ref{T8}) takes the form
\begin{equation}
{\displaystyle \sum_{k}}\left(x_{nk}p_{kn'}-p_{n'k}x_{kn}\right)=i\hbar\delta_{nn\text{\textasciiacute}}.\label{T12}
\end{equation}

The coefficients $x_{nk}$ and $p_{nk}$ can therefore be organized
as the elements of matrices $\hat{x}$ and $\hat{p}$, respectively,
with as many rows and columns as there are different states, and such
that Eq. (\ref{T12}) becomes
\begin{equation}
\left[\hat{x},\hat{p}\right]_{nn'}=i\hbar\delta_{nn'},\label{T14}
\end{equation}
which is precisely the matrix formula for the quantum commutator
\begin{equation}
\left[\hat{x},\hat{p}\right]=i\hbar.\label{T16}
\end{equation}

The (basic) quantum commutator is thus identified with \emph{the Poisson
bracket of the system's response variables $x,p$ with respect to
the normal field variables} $\left\{ a,a^{*}\right\} $. The matrix
coefficient $x_{nk}$ represents the (dipolar) response amplitude
of the system in state $n$ to the field mode $\left(nk\right)$.
It is clear that if the particle responds to a given mode $\left(nk\right)$
of the \noun{zpf}, it also has the capacity to respond to the mode
$\left(nk\right)$ of an external field. Whenever $x_{nk}\neq0$,
the system can thus make a transition from state $n$ to state $k$
and vice versa---as expressed in the well-known selection rules for
dipolar transitions. 

From this point on, using the fact that $\left[\hat{x},\hat{p}\right]$
defines a Lie algebra that satisfies the Leibniz rule and the Jacobi
identity, it is a direct matter to derive the expressions that complete
the quantum matrix formalism \cite{QSMF20}. 

To get a coherent picture of the physics behind the formalism it is
important to bear in mind at all times the meaning just disclosed
of the operators as response functions acting on the (state) vectors
of a Hilbert space. Thus for instance, by introducing the vectors
(summation over the repeated index $k$ is understood)
\begin{equation}
\boldsymbol{u}_{n}=x_{nk}\mathbf{\boldsymbol{\epsilon}}_{k},\;\boldsymbol{v}_{n}=p_{nk}\mathbf{\boldsymbol{\epsilon}}_{k},\label{T18}
\end{equation}
with $\mathbf{\boldsymbol{\epsilon}}_{k}$ unit orthogonal vectors,
and applying the Cauchy-Schwarz inequality to their scalar products,
\begin{equation}
\left|\boldsymbol{u}_{n}\cdot\boldsymbol{u}_{n}^{*}\right|\left|\boldsymbol{v}_{n}\cdot\boldsymbol{v}_{n}^{*}\right|\geq\left|\boldsymbol{u}_{n}\cdot\boldsymbol{v}_{n}^{*}\right|^{2},\label{T20}
\end{equation}
we obtain, with $p_{nk}=-im\omega_{kn}x_{nk}$,
\begin{equation}
\sum_{k\neq n}\left|x_{nk}\right|^{2}\sum_{k'\neq n}\left|p_{nk'}\right|^{2}\geq\left|m\sum_{k\neq n}\omega_{kn}\left|x_{nk}\right|^{2}\right|^{2}=\left(\frac{\hbar}{2}\right)^{2}.\label{T22}
\end{equation}
In writing the last equality we have used the Thomas-Reiche-Kuhn sum
rule, Eq. (\ref{T6}). This result may be written in the more familiar
form, using (\ref{T16}),
\begin{equation}
\left(\Delta\hat{x}\right)^{2}\left(\Delta\hat{p}\right)^{2}\geq\left|\left[\hat{x},\hat{p}\right]\right|^{2}=\left(\frac{\hbar}{2}\right)^{2},\label{T24}
\end{equation}
valid for any $n$. The Heisenberg inequality represents therefore,
from this perspective, a restriction on the minimum value of the product
of the entire set of response coefficients, with Planck's constant
representing the scale of the fluctuations of the \noun{zpf}. Further,
from the derivation just presented it is clear that the operators
$\hat{x},\hat{p}$ do not represent trajectories and there need not
be a phase-space description associated with them. 

\subsection{Statistical dynamic description in the Markov approximation\label{DYN}}

Let us now turn to the (initial) dynamics of the \noun{sed} particle,
which is usually described by the (non relativistic) equation of motion,
known as Braffort-Marshall equation 
\begin{equation}
m\ddot{\boldsymbol{x}}=\boldsymbol{f}(\boldsymbol{x})+m\tau\boldsymbol{\dddot{x}}+e\boldsymbol{E}(t),\label{C8}
\end{equation}
where $\boldsymbol{f}(\boldsymbol{x})$ is the external binding force
and $m\tau\boldsymbol{\dddot{x}}$ stands for the radiation reaction
force, with $\tau=2e^{2}/3mc^{3}$ \cite{TQD}. For an electron, $\tau\approx10^{-23}$
s. $\boldsymbol{E}(t)$ represents the electric component of the \noun{zpf}
taken in the long-wavelength approximation, with time correlation
given in the continuum limit by ($j,k=1,2,3$) \begin{subequations}
\label{EE}
\begin{equation}
\left\langle E_{k}(s)E_{j}(t)\right\rangle =\delta_{kj}\varphi(t-s),\label{C9a}
\end{equation}
 where the spectral function
\begin{equation}
\varphi(t-s)=\frac{2\hbar}{3\pi c^{3}}\intop_{0}^{\infty}d\omega\,\omega^{3}\cos\omega(t-s),\label{C9b}
\end{equation}
\end{subequations} corresponding to an energy $\hbar\omega/2$ per
mode, represents a highly colored noise. In order to analyze the effect
of the different forces on the particle dynamics, we introduce an
expansion of $\boldsymbol{x}(t)$ in terms of powers of $e$ 
\begin{equation}
\boldsymbol{\boldsymbol{\boldsymbol{x}}=\boldsymbol{\boldsymbol{x}^{\mathrm{(0})}+\boldsymbol{x}^{\mathrm{(1)}}+\boldsymbol{x}^{\mathrm{(2)}}+\ldots}}=\boldsymbol{\boldsymbol{x}}^{(0)}+\sum_{s=1}\boldsymbol{\boldsymbol{x}}^{(s)},\label{C12}
\end{equation}
where $x_{k}^{(s)}$ stands for the contribution of order $e^{s}$,
$e$ being here the coupling factor of the particle to the field.
(For neutral electromagnetic particles a different coupling needs
to be considered, which would depend on the charge distribution).
From Eq. (\ref{C8}) follows the hierarchy (summation over repeated
indices is understood)
\begin{equation}
m\ddot{x}_{i}^{(0)}=f_{i}(x^{(0)})+m\tau\dddot{x_{i}}^{(0)},\label{C16a}
\end{equation}

\begin{equation}
m\ddot{x}_{i}^{(1)}=\left.\frac{\partial f_{i}}{\partial x_{j}}\right|_{x^{(0)}}x_{j}^{(1)}+eE_{i}(t),\label{C16b}
\end{equation}

\begin{equation}
m\ddot{x}_{i}^{(2)}-\left.\frac{\partial f_{i}}{\partial x_{j}}\right|_{x^{(0)}}x_{j}^{(2)}=\frac{1}{2}\left.\frac{\partial^{2}f_{i}}{\partial x_{j}\partial x_{k}}\right|_{x^{(0)}}x_{j}^{(1)}x_{k}^{(1)},\label{C16c}
\end{equation}

\[
......
\]
By replacing, as is customary, the third-order time derivative by
its approximate first-order expression $\tau$$\left(df_{i}/dt\right)$
in Eq. (\ref{C16a}), one may readily see that the deterministic solution
of the homogeneous part of (\ref{C8}) (i. e., in the absence of the
\noun{zpf)} decays within a time lapse of the order of the lifetime
$\tau_{d}$ determined by the value of $\left|df_{i}/dt\right|$,
as was mentioned above. Further, from Eq. (\ref{C16b}) it follows
that $x_{i}^{(1)}$ is a purely stochastic variable, describing a
non-decaying motion driven by the electric component of the \noun{zpf},
which may be written in the form
\begin{equation}
x_{i}^{(1)}=e\int_{-\infty}^{t}ds\mathcal{\mathit{\mathcal{G_{\mathit{ik}}\mathit{\mathrm{(}t,s\mathrm{)\mathit{E_{k}(s),}}}}}}\label{C18}
\end{equation}
where the Green function $\mathcal{G}_{\mathit{ij}}\mathit{\mathrm{(}t,s\mathrm{)}}$
is a solution of the equation
\begin{equation}
m\frac{\partial^{2}}{\partial t^{2}}\mathcal{G}_{\mathit{ik}}\mathit{\mathrm{(}t,s\mathrm{)}}=\left.\frac{\partial f_{i}}{\partial x_{l}}\right|_{x^{(0)}}\mathcal{G}_{\mathit{lk}}\mathit{\mathrm{(}t,s\mathrm{),}}\label{C22a}
\end{equation}
with initial conditions
\begin{equation}
\mathcal{G}_{\mathit{ik}}\mathit{\mathrm{(}t,t)}=0,\qquad\mathrm{lim}{}_{s\rightarrow t}\frac{\partial}{\partial t}\mathcal{G}_{\mathit{ik}}\mathrm{(}t,s\mathrm{)=\mathit{\frac{\mathrm{1}}{m}\delta_{ik}.}}\label{C22b}
\end{equation}
The solution of Eq. (\ref{C22a}) satisfying these conditions is 
\begin{equation}
\mathcal{G}_{\mathit{ik}}\mathrm{(}t,s)=\left.\frac{\partial x_{i}(t)}{\partial p_{k}(s)}\right|_{x^{(0)}},\label{C23b}
\end{equation}
whence (\ref{C18}) becomes \begin{subequations} \label{xipi}
\begin{equation}
x_{i}^{(1)}=e\int_{-\infty}^{t}ds\left.\frac{\partial x_{i}(t)}{\partial p_{k}(s)}\right|_{x^{(0)}}\mathit{E_{k}(s),}\label{C26a}
\end{equation}
and its time derivative gives
\begin{equation}
p_{i}^{(1)}=e\int_{-\infty}^{t}ds\left.\frac{\partial p_{i}(t)}{\partial p_{k}(s)}\right|_{x^{(0)}}\mathit{E_{k}(s).}\label{C26b}
\end{equation}
\end{subequations} Notice that in these expressions, the derivatives
$\partial x_{i}(t)/\partial p_{k}(s)$, $\partial p_{i}(t)/\partial p_{k}(s)$,
which are functions of the exact solution of Eq. (\ref{C8}), are
to be calculated to zero order in $e$, i. e., at $\boldsymbol{x}^{(0)}$.
As is clear from Eq. (\ref{C16c}) and the following equations of
the hierarchy, the higher-order solutions $x_{i}^{(r)}$ with $r>1$
are all determined (although indirectly) by the field.

These results describe with revealing detail several important aspects
of the dynamics. Equations (\ref{xipi}) show how $x_{i}^{(1)}(t)$
and $p_{i}^{(1)}(t)$ are constructed by the \noun{zpf} starting from
zero, at specific rates determined by the external force. We observe
that along the evolution of the system, the dynamics undergoes a qualitative
change, the \noun{zpf} moving from being merely the source of some
noise impressed on the particle motion, to injecting upon it indeterministic
properties and gaining control of the response, after a period of
the order of $\tau_{d}$.

Alternatively, a statistical treatment of the problem, following an
approach that is standard in the theory of stochastic processes, starts
from Eq. (\ref{C8}) and leads by means of a smoothing process to
a generalized Fokker-Planck equation (\noun{gfpe}) for the probability
density $Q(\boldsymbol{x},\boldsymbol{p},t)$ (\cite{Rice}, \cite{Kam 81}
p. 209). The \noun{gfpe} is an integro-differential equation, or equivalently,
a differential equation containing an infinite number of time-dependent
terms, which express the memory-laden buildup of the diffusion. At
the initial time the system is far from equilibrium and there is no
diffusion; once the Markov approximation applies, the FPE is reduced
to a true Fokker-Planck equation \cite{Prabhu}, i. e., to a second-order
differential equation, 
\begin{equation}
\frac{\partial Q}{\partial t}+\frac{1}{m}\frac{\partial}{\partial x_{i}}p_{i}Q+\frac{\partial}{\partial p_{i}}f_{i}Q+m\tau\frac{\partial}{\partial p_{i}}\dddot{x}_{i}Q=D_{ij}^{px}\frac{\partial^{2}Q}{\partial p_{i}\partial x_{j}}+D_{ij}^{pp}\frac{\partial^{2}Q}{\partial p_{i}\partial p_{j}},\label{C28}
\end{equation}
with diffusion tensors given by (\cite{TEQ} Ch.4 and refs. therein)
\begin{subequations} \label{dif tens}

\begin{equation}
D_{ij}^{px}(t)=e\left\langle x_{i}E_{j}\right\rangle =e^{2}\int_{-\infty}^{t}ds\left.\frac{\partial x_{i}(t)}{\partial p_{k}(s)}\right|_{x^{(0)}}\left\langle E_{k}(s)E_{j}(t)\right\rangle ,\label{C10a}
\end{equation}
\begin{equation}
D_{ij}^{pp}(t)=e\left\langle p_{i}E_{j}\right\rangle =e^{2}\int_{-\infty}^{t}ds\left.\frac{\partial p_{i}(t)}{\partial p_{k}(s)}\right|_{x^{(0)}}\left\langle E_{k}(s)E_{j}(t)\right\rangle ,\label{C10b}
\end{equation}
\end{subequations} where the brackets denote averaging over the realizations
of the stochastic field, and the derivatives are calculated at $\boldsymbol{x}^{(0)}$. 

Notice that by taking the product of $E_{j}(t)$ with $x_{i}^{(1)}$
given by Eq. (\ref{C26a}) and averaging, one obtains precisely Eq.
(\ref{C10a}), and similarly, the averaged product of $E_{j}(t)$
with $p_{i}^{(1)}$ given by Eq. (\ref{C26b}) gives Eq. (\ref{C10b}).
This shows that the Markov approximation is consistent with the notion
that the field has taken control of the variables $x,p$. In the radiationless
approximation, the \noun{fpe} (\ref{C28}) leads to the Schrödinger
equation in configuration space (\cite{TEQ} Ch. 4). Further, since
$e\left\langle \boldsymbol{x}\cdot\boldsymbol{E}\right\rangle $ is
the mean work realized by the field on the particles, $\mathrm{Tr}D^{px}$
represents a radiative contribution to the mean energy of the particles.
Indeed, this has been shown to correspond to the nonrelativistic formula
for the Lamb shift \cite{TQD,RMF76}, which assigns a clear meaning
to the origin of the Lamb shift, well in accord with the intuitive
interpretation suggested originally by Welton \cite{Welt48} (detailed
discussions of this point can be seen in \cite{TEQ} Ch. 6, and \cite{Mil19}
Ch. 7). Similarly, $\mathrm{Tr}D^{pp}$ obtained from Eq. (\ref{C10b})
is related with the mean power absorbed from the field along the orbital
motion; we shall deal with this term in subsection \ref{EB}.

Since, as said above, the variables $x,p$ have become controlled
by the field, we may introduce into the above expressions the results
obtained for the kinematics in subsection \ref{KIN}. We observe that
in Eqs. (\ref{xipi}) and (\ref{dif tens}), the integrands contain
factors of the form $\partial x_{i}(t)/\partial p_{j}(s)$, $\partial p_{i}(t)/\partial p_{j}(s)$,
which may be written in terms of Poisson brackets,
\begin{equation}
\partial x_{i}(t)/\partial p_{k}(s)=\left\{ x_{k}(s),x_{i}(t)\right\} _{xp},\;\partial p_{i}(t)/\partial p_{k}(s)=\left\{ x_{k}(s),p_{i}(t)\right\} _{xp},\label{C32}
\end{equation}
calculated at $\boldsymbol{x}^{(0)}$, i. e., to zero order in $e.$
According to what we have learned in subsection 3.2, for times $s,t$
larger than $\tau_{d}$ these Poisson brackets should be replaced
by the corresponding quantum commutators, to be calculated in a given
stationary state, say $n$ (for ease of notation we omit the subindex
$n$ when possible),
\begin{equation}
\left\{ x_{k}(s),x_{i}(t)\right\} _{xp}\rightarrow\frac{1}{i\hbar}\left[\hat{x}_{k}(s),\hat{x}_{i}(t)\right],\;\left\{ x_{k}(s),p_{i}(t)\right\} _{xp}\rightarrow\frac{1}{i\hbar}\left[\hat{x}_{k}(s),\hat{p}_{i}(t)\right].\label{C34}
\end{equation}
 Therefore, by setting $t_{o}=-\infty$ we may safely neglect the
initial contribution to the integral for $t_{o}\leq s\leq t_{o}+\tau_{d}$
and write Eqs. (\ref{xipi}) in the form\begin{subequations} \label{xipi^1}
\begin{equation}
x_{i}^{(1)}=\frac{e}{i\hbar}\int_{-\infty}^{t}ds\left[\hat{x}_{k}(s),\hat{x}_{i}(t)\right]\mathit{E_{k}(s),}\label{C36a}
\end{equation}
\begin{equation}
p_{i}^{(1)}=\frac{e}{i\hbar}\int_{-\infty}^{t}ds\left[\hat{x}_{k}(s),\hat{p}_{i}(t)\right]\mathit{E_{k}(s).}\label{C36b}
\end{equation}
\end{subequations} Similarly, from Eqs. (\ref{dif tens}) we obtain
for the diffusion tensors \begin{subequations} \label{dif tens^1}
\begin{equation}
D_{ij}^{px}(t)=e\left\langle x_{i}E_{j}\right\rangle =\frac{e^{2}}{i\hbar}\int_{-\infty}^{t}ds\left[\hat{x}_{k}(s),\hat{x}_{i}(t)\right]\left\langle E_{k}(s)E_{j}(t)\right\rangle ,\label{C38a}
\end{equation}
\begin{equation}
D_{ij}^{pp}(t)=e\left\langle p_{i}E_{j}\right\rangle =\frac{e^{2}}{i\hbar}\int_{-\infty}^{t}ds\left[\hat{x}_{k}(s),\hat{p}_{i}(t)\right]\left\langle E_{k}(s)E_{j}(t)\right\rangle .\label{C38b}
\end{equation}
\end{subequations} Using Eqs. (\ref{EE}) for the time correlation
of the \noun{zpf}, we get \begin{subequations} \label{dif tens^1-1}
\begin{equation}
\mathrm{Tr}D^{px}(t)=e\left\langle \boldsymbol{x}\cdot\boldsymbol{E}\right\rangle =-\frac{2ie^{2}}{3\pi c^{3}}\int_{-\infty}^{t}ds\left[\hat{x}_{i}(s),\hat{x}_{i}(t)\right]\intop_{0}^{\infty}d\omega\,\omega^{3}\cos\omega(t-s),\label{C40a}
\end{equation}
\begin{equation}
\mathrm{Tr}D^{pp}(t)=e\left\langle \boldsymbol{p}\cdot\boldsymbol{E}\right\rangle =-\frac{2ie^{2}}{3\pi c^{3}}\int_{-\infty}^{t}ds\left[\hat{x}_{i}(s),\hat{p}_{i}(t)\right]\intop_{0}^{\infty}d\omega\,\omega^{3}\cos\omega(t-s).\label{C40b}
\end{equation}
\end{subequations} 

\subsection{Energy balance \label{EB}}

We now use the above result, Eq. (\ref{C40b}), to analyze the energy
balance for the system in a state $n$. The equation of evolution
for the average energy is obtained by multiplying the dynamical equation
(\ref{C8}) with $\boldsymbol{p}$ and averaging over the ensemble,
\begin{equation}
\frac{d}{dt}\left\langle H\right\rangle _{n}=\tau\left\langle \boldsymbol{p}\cdot\boldsymbol{\dddot{x}}\right\rangle _{n}+\frac{e}{m}\left\langle \boldsymbol{p\cdot}\boldsymbol{E}(t)\right\rangle _{n}.\label{EB2}
\end{equation}
The second term on the rhs represents the average power absorbed from
the field. Writing the commutator in (\ref{C40b}) in terms of the
matrix elements corresponding to state $n$ with the help of Eqs.
(\ref{T4}) and (\ref{T10}) gives
\begin{equation}
\left[\hat{x}_{i}(s),\hat{p_{i}}(t)\right]_{nn}=2im\sum_{i}\sum_{k}\left|x_{ink}\right|^{2}\omega_{ikn}\cos\omega_{ikn}(t-s),\label{EB4}
\end{equation}
whence
\[
\frac{}{}\left(\mathrm{Tr}D^{pp}\right)_{n}=e\left\langle \boldsymbol{p}\cdot\boldsymbol{E}\right\rangle _{n}
\]
\[
=\frac{4me^{2}}{3\pi c^{3}}\sum_{i}\sum_{k}\left|x_{ink}\right|^{2}\omega_{ikn}\intop_{0}^{\infty}d\omega\,\omega^{3}\int_{-\infty}^{t}ds\cos\omega(t-s)\cos\omega_{ikn}(t-s)
\]
\begin{equation}
=\frac{2me^{2}}{3c^{3}}\sum_{i}\sum_{k}\left|x_{ink}\right|^{2}\omega_{ikn}^{4}\left[\delta(\omega-\omega_{ikn})-\delta(\omega+\omega_{ikn})\right].\label{EB6}
\end{equation}
In its turn, the first term on the rhs of (\ref{EB2}) represents
the average power lost by radiation reaction, which is readily calculated
using Eqs. (\ref{T4}),
\begin{equation}
\tau\left(\boldsymbol{p}\cdot\boldsymbol{\dddot{x}}\right)_{n}=-\frac{2e^{2}}{3c^{3}}\sum_{i}\sum_{k}\left|x_{ink}\right|^{2}\omega_{ikn}^{4}.\label{EB8}
\end{equation}
Introducing Eqs. (\ref{EB6}) and (\ref{EB8}) in (\ref{EB2}) one
gets
\begin{equation}
\frac{d}{dt}\left\langle H\right\rangle _{n}=-\frac{4e^{2}}{3c^{3}}\sum_{i}\sum_{k}^{\omega_{ink}>0}\left|x_{ink}\right|^{2}\omega_{ikn}^{4}.\label{EB10}
\end{equation}
When the system is in its ground state there is no contribution to
the sum in Eq. (\ref{EB10}), which confirms that the energy lost
by radiation is compensated in the mean by the energy extracted from
the \noun{zpf} and detailed energy balance holds. By contrast, when
the system is in an excited state, Eq. (\ref{EB10}) gives for the
rate of change of the energy (\cite{TEQ}, Ch. 6) \begin{subequations}\label{Ank}
\begin{equation}
\frac{d}{dt}\left\langle H\right\rangle _{n}=-\sum_{i}\sum_{k}^{\omega_{ink}>0}\hbar\omega_{ink}A_{ink},\label{EB12a}
\end{equation}
with
\begin{equation}
A_{ink}=\frac{4e^{2}}{3\hbar c^{3}}\sum_{i}\left|x_{ink}\right|^{2}\omega_{ikn}^{3},\label{EB12b}
\end{equation}
\end{subequations} which coincides with the \noun{qed} formula for
the Einstein spontaneous emission coefficient and serves to demonstrate
that the radiation reaction and the \noun{zpf} contribute equal parts
to the spontaneous emission rate (see \cite{Mil19} for a discussion
of this point in the context of \noun{qed}).

\section{Final comments and conclusions}

To recap, firstly we have introduced the electromagnetic \noun{zpf}
as part of the quantum ontology with the purpose of addressing several
longstanding issues of quantum mechanics. In particular, its presence
assigns a physical cause to the quantum fluctuations and thus accounts
for the indeterministic nature of the quantum phenomenon. By introducing
the \noun{zpf} we are considering that the (quantum) world is made
of electric charges (or electromagnetic particles, more generally)
that fill the space with radiation as a result of their permanent
wiggling. This does not preclude the possibility that other fields
should be required at some point to complete the quantum ontology;
to arrive at \noun{qm}, however, the electromagnetic vacuum proves
to be sufficient. The image of an isolated atom or a quantum particle
in empty space appears thus as an idealization with no real counterpart.
If it were experimentally feasible to introduce an atom or any quantum
system in a volume completely free of radiation, including the \noun{zpf},
this would provide a valuable possibility to subject the theory to
testing.

Secondly, the introduction of the \noun{zpf} is seen to represent
more than the mere addition of a new element to the ontology. It changes
the (apparently) mechanical \textit{nature} of the quantum problem
to an electrodynamic one. In addition to ensuring (and explaining)
the atomic stability by compensating for the energy lost by radiation,
it offers a possible explanation for the so-called quantum jumps \cite{qjumps},
along with an understanding of the mechanism of entanglement \cite{Ent}
and of the electron spin as an emergent property \cite{spin}, as
well as the (nonrelativistic) radiative corrections proper of \noun{qed}
\cite{TEQ}.

Thirdly, with the present work we have completed the picture by showing
that as a consequence of dissipation, after a time lapse of order
$\tau_{d}$ the \noun{zpf} becomes the driving force for the particle\noun{.}
We embody this central role of the \noun{zpf} in a \textit{kinematic
condition}\textit{\emph{,}} which explains what represents perhaps
the most obscure of all quantum features: the representation of dynamical
variables by operators in a Hilbert space, the elements of which are
the response functions associated with the (dipolar) transitions induced
by the field. This, in its turn, helps clarify another intriguing
question of \noun{qm} that sounds as an oxymoron: how is it possible
that the description provided by the quantum formalism of a \textit{stationary}
quantum state is to be made in terms of a collection of \textit{transition}
amplitudes between states? The suggested functionality of the transition
coefficients as the building blocks of the operator representation
has moreover an historical value, since in Born's hands (and as was
involuntarily suggested by Heisenberg) they came to be the omnipresent
elements of the quantum description.

A point of contention is whether the resulting theory is still ``classical''.
The inclusion of the \noun{zpf} can be seen as either a \emph{relinquishment}
of classical physics (by the introduction of a foreign element), an
\emph{addition,} or even a \emph{correction} to classical physics
(by recognizing the reality of this field). These diverse points of
view are present in the literature; see in particular Ref. \cite{Khrn07}.
That the inclusion of the \noun{zpf} represents a rupture with classical
physics is confirmed, some would say, by the fact that it eventually
leads to quantum mechanics, which is no more classical physics. Interestingly,
hardly anyone would argue that Brownian motion is no more ``classical
physics''---even though the chaotic and diffusive behavior of the
molecules cannot be explained without consideration of the embedding
medium. In fact, classical physics began to consider stochastic processes
as part of it with the advent of Brownian motion in the hands of Einstein,
followed by Smoluchowski. Thus from the point of view of \noun{sed,}
one might consider Einstein the grandfather of \noun{qm}---and one
of its (pre-)founders. There are of course elements that make the
quantum stochastic process differ \emph{essentially} from Brownian
motion, the central one being the temporal and spatial correlations
of the \noun{zpf }as opposed to a white noise. The commutator $\left[\hat{x},\hat{p}\right]$
has a universal (state-independent) form thanks to the functional
dependence on $\omega$ in the transformation equations (\ref{K22}).
Further, the memory associated with the colored noise spectrum is
essential for the buildup of diffusion during the transition to the
quantum regime. The question then arises of whether in other cases
where there is a correlated, wavelike, stationary background noise,
one should expect quantization to emerge; this is an issue that has
prompted a whole new branch of research under the name of hydrodynamic
quantum analogies \cite{BushOza21}.

From the perspective gained with the present work, one may say that
quantum dynamics is a brand-new variant and extension of classical
physics into the stochastic domain, which finds a place of its own,
both because of the distinctive behavior of quantum systems and for
the wealth of phenomena and applications to which it gives rise.

The authors are grateful to Andrea Valdés-Hernández for her critical
revision and valuable comments to the manuscript.

\end{document}